\documentclass[showkeys,twocolumn,showpacs,preprintnumbers,amsmath,amssymb,prl,superscriptaddress]{revtex4-1}
\usepackage{graphicx}
\usepackage{dcolumn}
\usepackage{bm}
\usepackage{latexsym,epsfig}
\usepackage{hyperref}
\usepackage{here}
\usepackage{siunitx}

\newcommand\mg{Mg$^+$}
\newcommand\mgh{MgH$^+$}
\newcommand\gs{$X ^1\Sigma^+$}
\newcommand\as{$A ^1\Sigma^+$}
\newcommand\cs{$C ^1\Sigma^+$}
\newcommand\zz{$|\nu$=$0,J$=$0\rangle_X$}

\newcommand\zt{$|\nu$=$0,J$=$2\rangle_X$}
\newcommand\oo{$|\nu$=$1,J$=$1\rangle_X$}

\begin{document}

\preprint{\today}

\title{Decay rate measurement of the first vibrationally excited state of MgH$^+$ \\ in a cryogenic Paul trap}
\vspace{0.5cm}
\author{O.~O. Versolato}
\email{oscar.versolato@mpi-hd.mpg.de}
\author{M. Schwarz}
\affiliation{Max-Planck-Institut f\"ur Kernphysik, Saupfercheckweg 1, 69117 Heidelberg, Germany}
\author{A. K. Hansen}
\author{A. D. Gingell}
\affiliation{QUANTOP - The Danish National Research Foundation Center for Quantum Optics, Department of Physics and Astronomy, Aarhus University, DK-8000 
Aarhus C, Denmark}
\author{A. Windberger}
\affiliation{Max-Planck-Institut f\"ur Kernphysik, Saupfercheckweg 1, 69117 Heidelberg, Germany}
\author{\\ {\L}. K{\l}osowski}
\affiliation{Institute of Physics, Faculty of Physics, Astronomy and Informatics, Nicolaus Copernicus University, Grudziadzka 5, 87-100 Torun, Poland}
\author{J. Ullrich}
\affiliation{Max-Planck-Institut f\"ur Kernphysik, Saupfercheckweg 1, 69117 Heidelberg, Germany}
\affiliation{Physikalisch-Technische Bundesanstalt, Bundesallee 100, 38116, Braunschweig, Germany}
\author{F. Jensen}
\affiliation{Department of Chemistry, Aarhus University, DK-8000 
Aarhus C, Denmark}
\author{J.~ R.~\surname{Crespo~L\'opez-Urrutia}}
\affiliation{Max-Planck-Institut f\"ur Kernphysik, Saupfercheckweg 1, 69117 Heidelberg, Germany}
\author{M. Drewsen}
\email{drewsen@phys.au.dk}
\affiliation{QUANTOP - The Danish National Research Foundation Center for Quantum Optics, Department of Physics and Astronomy, Aarhus University, DK-8000 
Aarhus C, Denmark}

\begin{abstract}
\noindent 
We present a method to measure the decay rate of the first excited vibrational state of simple polar molecular ions being part of a Coulomb crystal in a cryogenic linear Paul trap. Specifically, we have monitored the decay of the $|\nu$=$1,J$=$1 \rangle_X$ towards the $|\nu$=$0,J$=$0 \rangle_X$ level in \mgh\ by saturated laser excitation of the $|\nu$=$0,J$=$2 \rangle_X$-$|\nu$=$1,J$=$1 \rangle_X$ transition followed by state selective resonance enhanced two-photon dissociation out of the $|\nu$=$0,J$=$2 \rangle_X$ level. The technique enables the determination of decay rates, and thus absorption strengths, with an accuracy at the few percent level.
\end{abstract}
 \keywords{}

\maketitle
Diatomic hydrides are the simplest molecules with dipole-allowed rovibrational transitions. Such species like CH, OH, and NH can be readily observed in the interstellar medium (ISM) both in their neutral and singly-ionized states \cite{weinreb1963radio,singh1987one,wardle2002supernova}. In general, their abundances in the ISM are derived from mm-wave and deep infra-red (IR) absorption and emission lines \cite{snow2008ion} for which precise knowledge of the relevant rovibrational transition strengths is required. The determination of the decay rates of single rovibrationally excited states represents the most direct way of determining these strengths \cite{Erman1977,Meijer2005}. However, a scarcity of accurate laboratory data exist since lifetimes ranging up to many seconds require both excellent vacuum conditions and suppression of \mbox{black-body radiation (BBR).}

In this Letter, we present a measurement of the decay of the $|\nu$=$1,J$=$1 \rangle_X$ towards the $|\nu$=$0,J$=$0 \rangle_X$ level in \mgh\, by saturated laser excitation of the \mbox{$|\nu$=$0,J$=$2 \rangle_X$-$|\nu$=$1,J$=$1 \rangle_X$} transition followed by rotational-state-selective two-color 1+1 resonance-enhanced multi-photon dissociation (REMPD) out of the $|\nu$=$0,J$=$2 \rangle_X$ level \cite{DrewsenNature2010}. Here, $|\nu,J \rangle_X$ denotes the level with vibrational quantum number $\nu$ and rotational quantum number $J$ in the electronic ground state \gs. The molecular ions were trapped in a recently commissioned cryogenic Paul trap, CryPTEx \cite{RSI2012}, operating at a temperature of $\sim$5\,K and providing excellent vacuum conditions. In this setting, the molecular ions were sympathetically cooled into a Coulomb crystal through Coulomb interaction with Doppler laser cooled Mg$^+$ \mbox{ions \cite{Molhave2000b}}.  

\begin{figure}[b]
\centering
\includegraphics[width=0.48\textwidth]{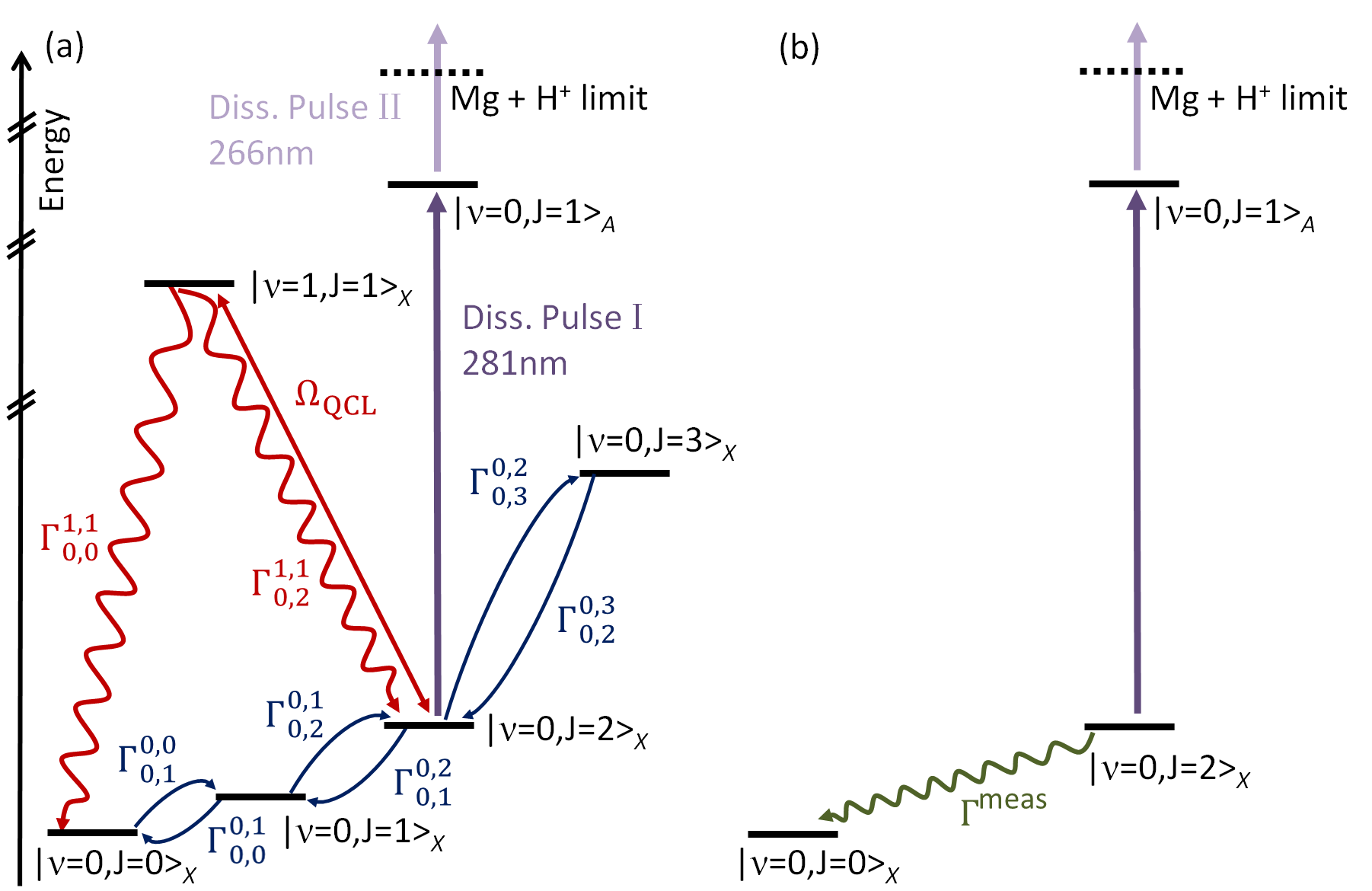}
\caption{Level scheme of \mgh\, depicting the relevant rovibrational levels $|\nu,J \rangle_{X}$ in the \gs\, electronic potential with indication of important transitions for the rovibrational dynamics. Also shown is the $|\nu$=$0,J$=$1 \rangle_A$ level of the \as\, electrical potential used in the REMPD state detection (see text for details).}
\label{fig:levelscheme}
\end{figure}
In Fig.\,\ref{fig:levelscheme}, a level scheme depicting the lowest relevant rovibrational levels $|\nu,J \rangle_{X}$ in the \gs\, potential of \mgh\, is presented. Transitions induced between the various states $|\nu',J' \rangle_{X}$ due to BBR and collisions with residual background gas particles are characterized by rate constants $\Gamma^{\nu,J}_{\nu',J'}$. For simplicity only dipole-allowed transitions are indicated. A linearly polarized Quantum Cascade Laser (QCL) operating at 6.2\,$\mu$m couples the \zt\, and \oo\, levels \cite{Balfour1973} at an effective Rabi frequency of $\Omega_\textrm{QCL}$, and enables the decay \oo$\rightarrow$\zz. As will be confirmed later, the cryogenic setup assures that $\Gamma^{0,J'}_{0,2}$, $\Gamma^{0,2}_{0,J'} \ll \Gamma^{1,1}_{0,0}$, and the intensity of the QCL is sufficient to achieve $\Gamma^{1,1}_{0,2}$, $\Gamma^{1,1}_{0,0} \ll \Omega_\textrm{QCL}$. Combined with the fact that the magnetic sub-states of the $|\nu$=$0,J$=$2 \rangle_X$ level mix at a rate $\Gamma_{B\textrm{-Mix.}} \gg \Gamma^{1,1}_{0,0}$ due to uncompensated magnetic fields (see later), it is possible to dramatically simplify the complex situation of Fig.\,\ref{fig:levelscheme} (a) to the one depicted in Fig.\,\ref{fig:levelscheme} (b) with the corresponding rate equation model governing the decay of the population $p_{0,2}$ \mbox{of the level \zt}
\begin{equation}\label{decay}
\dot p_{0,2} =- \frac{3}{8} \Gamma^{1,1}_{0,0} p_{0,2} \equiv - \Gamma^{\text{meas}} p_{0,2}.
\end{equation}
Here, the factor 3/8 arises from the fact that under the above conditions there will always be an equal population in all of the magnetic sub-states of the $|\nu$=$0,J$=$2 \rangle_X$ (5 sub-states) and $|\nu$=$1,J$=$1 \rangle_X$ (3 sub-states) levels, with only the latter sub-states decaying towards the $|\nu$=$0,J$=$0 \rangle_X$ level with equal rates $\Gamma^{1,1}_{0,0}$. Measurement of the population of the \zt\, level as function of time will hence result in a decay rate $\Gamma^{\text{meas}}$ from which the fundamental decay rate $\Gamma^{1,1}_{0,0}$ is readily obtained. 

REMPD of ions trapped in a Coulomb crystal provides for a particularly convenient means of measuring state populations \cite{DrewsenNature2010}. In the above dynamics, we have not considered vibrational excitations other than the one driven by the QCL, since excitations out of the vibrational ground state will happen at rates much slower than $\Gamma^{1,1}_{0,0}$ (a very conservative upper bound of the excitation rate due to BBR, collisions and off-resonant excitations by the cooling laser is 0.1\,s$^{-1}$). Theory predicts the rates $\Gamma^{1,1}_{0,0}$, $\Gamma^{1,1}_{0,2}$ to be of the order 5-10\,Hz.
\begin{figure}[b]
\centering
\includegraphics[width=0.48\textwidth]{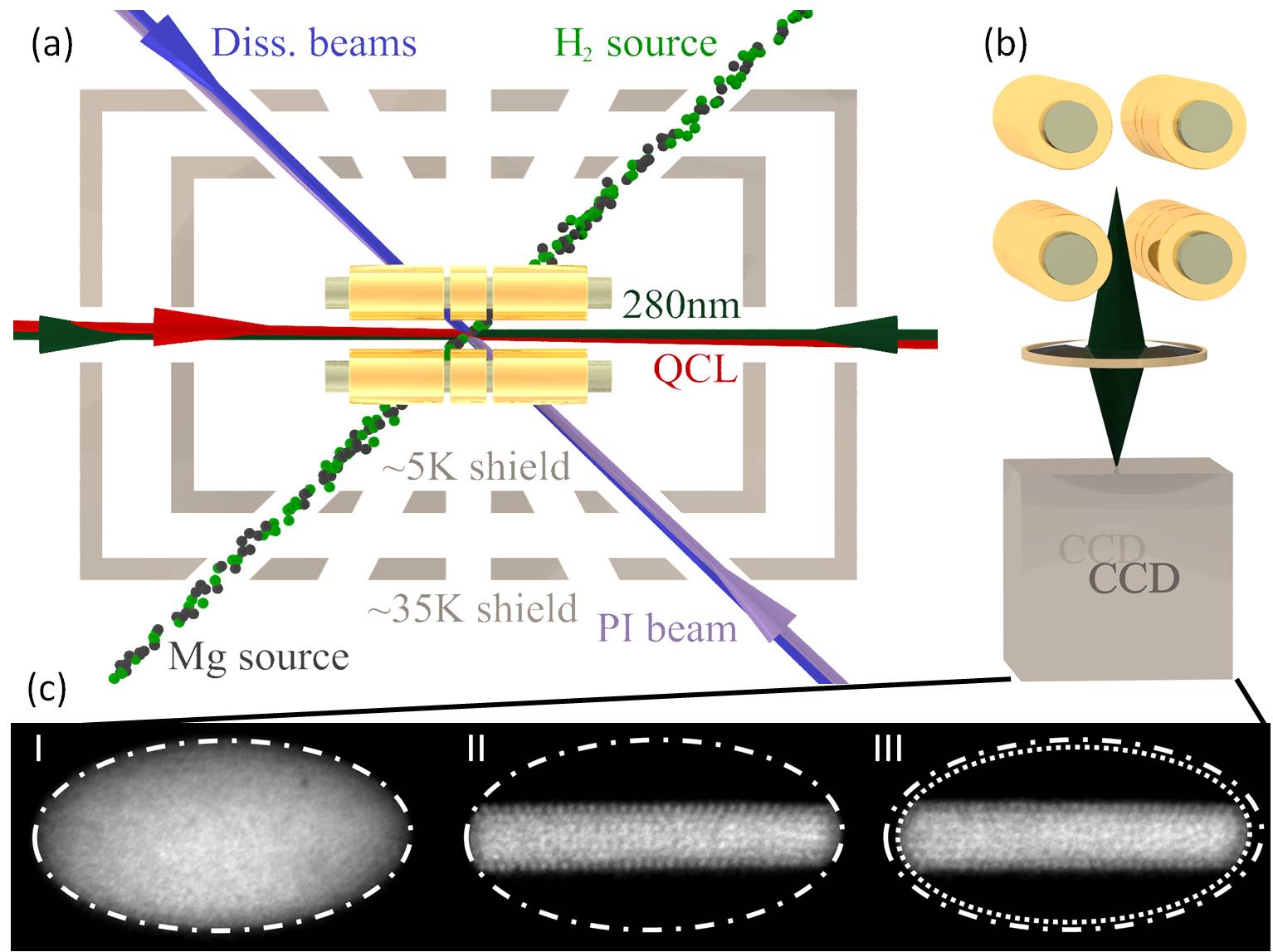}
\caption{(a) Schematics of the experimental setup: Cryogenic Paul trap electrodes (yellow), parts of the $\sim$5\,K inner and $\sim$35\,K outer shields (gray), laser beams for dissociation (purple), photoionization (lavender), and laser cooling at 280\,nm (dark green), the QCL beam (red), the Mg source (black spheres), and the H$_2$ source (green spheres). (b) Front view of the trap, including the imaging system consisting of a plano-convex spherical lens ($f$=40\,mm), and an image intensified CCD (charge-coupled device) camera (3.2$\times$ magnification) mounted below the trap. The collected fluorescence light creates a 2D projection image of the Mg$^+$ ions in the Coulomb crystals (c). (c) Images of Coulomb crystals during an experimental sequence: I, pure Mg$^+$ crystal consisting of 6630(200) ions (after photoionization loading). II, crystal composed of 1250(100) fluorescing Mg$^+$ ions and 5200(200) MgH$^+$ non-fluorescent ions (after reactions). III, after a pair of REMPD pulses addressing the P(2) transition of the resonant step was applied. The dashed-dotted ellipses indicate the outer boundary of the initial ion Coulomb crystal, while the dotted one indicates the reduced sized crystal after a set of REMPD pulses have been applied. All images correspond to a CCD exposure time of 500\,ms.}
\label{fig:setup}
\end{figure} 
To obtain the data needed to extract the $\Gamma^{1,1}_{0,0}$ decay rate, the following experimental sequence is repeated for various exposure times \mbox{of the QCL.}  

(i) First, $^{24}$Mg$^+$ ions are loaded into the trap via isotope-selective two-photon ionization of Mg atoms in an effusive beam \cite{kjaergaard2000,madsen2000} (see Fig.\,\ref{fig:setup} (a)). In the first step a 285\,nm laser crosses the atomic beam at right angles, and resonantly excites the \mbox{$3s^2$ $^1S_0$ - $3s3p$ $^1P_1$} transition of the Mg atom nearly Doppler-freely. The second photon leading to ionization is either yet another 285\,nm photon or a 280\,nm photon from one of the two counter-propagating laser beams providing the light for the Doppler cooling of the Mg$^+$ ions via the \mbox{$3s$ $^2S_{1/2}$ - $3p$ $^2P_{3/2}$} transition. The 285\,nm ionization beam is applied until a Coulomb crystal containing typically 5--7$\times10^3$ \mg\, ions is produced. The number of ions in the crystal is determined from two-dimensional projection images of the crystals resulting from recording fluorescence from the Mg$^+$ ions by an image-intensified CCD camera. Based on the fact that cold ions form spheroidal shaped crystals with constant ion densities, the total number of ions in these rotationally symmetric crystals can be easily deduced from the elliptical projection images (see \mbox{Fig.\,\ref{fig:setup} (c,I)}) \cite{madsen2000,hornekaer2001}. 

(ii) Next, the crystal is exposed to an effusive molecular beam of H$_2$ molecules, and \mgh\, molecules are formed through the reaction \mbox{$\text{Mg}^{+} (3p$ $^2P_{3/2}) + \text{H}_{2} \longrightarrow \text{MgH}^{+} + \text{H}$} \cite{PhysRevA.62.011401}. When about 80\% of the \mg\, ions in the crystal have reacted and the \mgh\, ions have been sympathetically cooled to become part of a Coulomb crystal, the needle valve controlling the H$_2$ flux is closed. In spite of this intermittent gas input, an H$_2$ background gas number density well below that equivalent to a pressure of 10$^{-11}$\,mbar (at 300\,K) is achieved after closing this valve. Even though the \mgh\, ions cannot directly be detected through fluorescence measurements, their abundance can be deduced indirectly by the presence of the fluorescence from remaining \mg\, ions \cite{PhysRevA.62.011401,hornekaer2001} (see \mbox{Fig.\,\ref{fig:setup} (c,II)}). 

(iii) After shutting off the H$_2$ molecular beam, the \mgh\, ions are left to equilibrate rovibrationally for 4 minutes before the QCL addressing the \zt-\oo\, transition irradiates the \mgh\, ions for a fixed time varying from a few ms to a \mbox{few s}. The QCL exposure time is controlled by employing a mechanical shutter with a 200\,$\mu$s rise time. 

(iv) Finally, at the end of the QCL exposure period, the population in the \zt\, level is determined by applying a pair of linearly polarized REMPD laser pulses. A 281\,nm pulse (8.5\,$\mu$J, 9\,ns, $\sim$0.5\,mm$^2$ spot size) resonantly addresses the \zt-$|\nu$=$0,J$=$1 \rangle_A$ transition whilst an overlapping 266\,nm pulse (1.1\,mJ, 9\,ns, $\sim$0.5\,mm$^2$ spot size) provides excitation to the dissociative \cs\, potential. It should be noted that the resonant P-branch excitation leads only to dissociation of the $M_J$=$\{-1,0,1\}$ sub-states. In \mbox{Fig.\,\ref{fig:setup} (c,III)}, the change in the size of the overall crystal is immediately observed, and a refined analysis shows that only a fraction of the \mgh\, ions have been lost, but no \mg\, ions. Since only a fraction of the molecules are dissociated (the ones in the \zt\, level), steps (iii) and (iv) are repeated until about half of the originally produced molecular ions are dissociated before returning to the first step (i). 
%
\begin{figure}
\includegraphics[width=87mm]{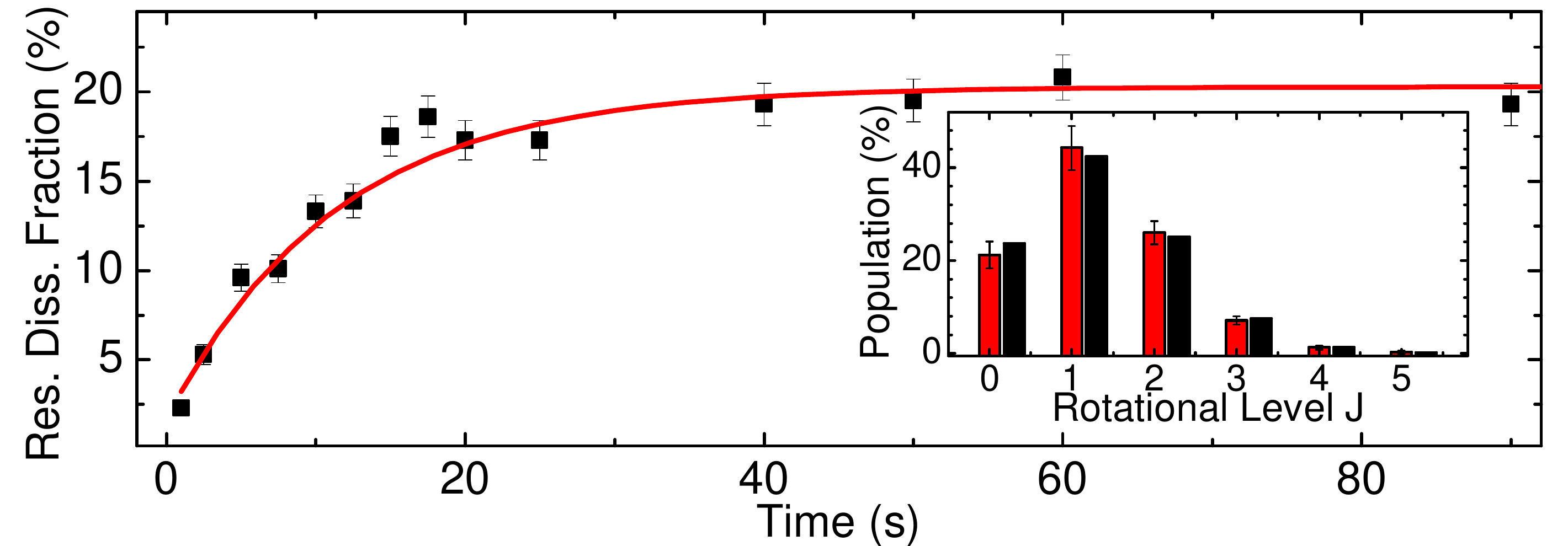}
\caption{Main figure: Refilling dynamics of the population of the \zt\, level obtained by measuring the resonantly dissociated molecular fraction after initial depletion versus time. The red curve represents an exponential fit to the data with rate 0.088(11)\,s$^{-1}$. Insert: measured equilibrium rotational state distribution depicted by red bars (1$\sigma$ standard deviations shown). The black bars display a (36\,K) thermal distribution for comparison.}
\label{fig:refilling_ensemble_v1}
\end{figure}
%
%
%
In Fig.\,\ref{fig:refilling_ensemble_v1} the equilibrium rotational state distribution constructed from REMPD measurements (for method, see \citep{DrewsenNature2010}) before the application of the QCL is presented (insert), together with a graph showing the refilling dynamics of the \zt\, level when initially emptied (main figure). Already after 1 minute the steady-state distribution has clearly been reestablished, indicating that the 4-minute waiting-time introduced in the experimental step (iii) should be more than sufficient. The fitted refilling rate of $0.088(11)$\,s$^{-1}$ (numbers in brackets indicate one standard deviation uncertainties) is furthermore sufficiently slow to ensure that the assumption $\Gamma^{0,J'}_{0,2}, \Gamma^{0,2}_{0,J'} \ll \Gamma^{1,1}_{0,0}$ required for the simplified decay governed by equation\,\eqref{decay} is valid under our experimental conditions. 

\begin{figure}[b]
\includegraphics[width=87mm]{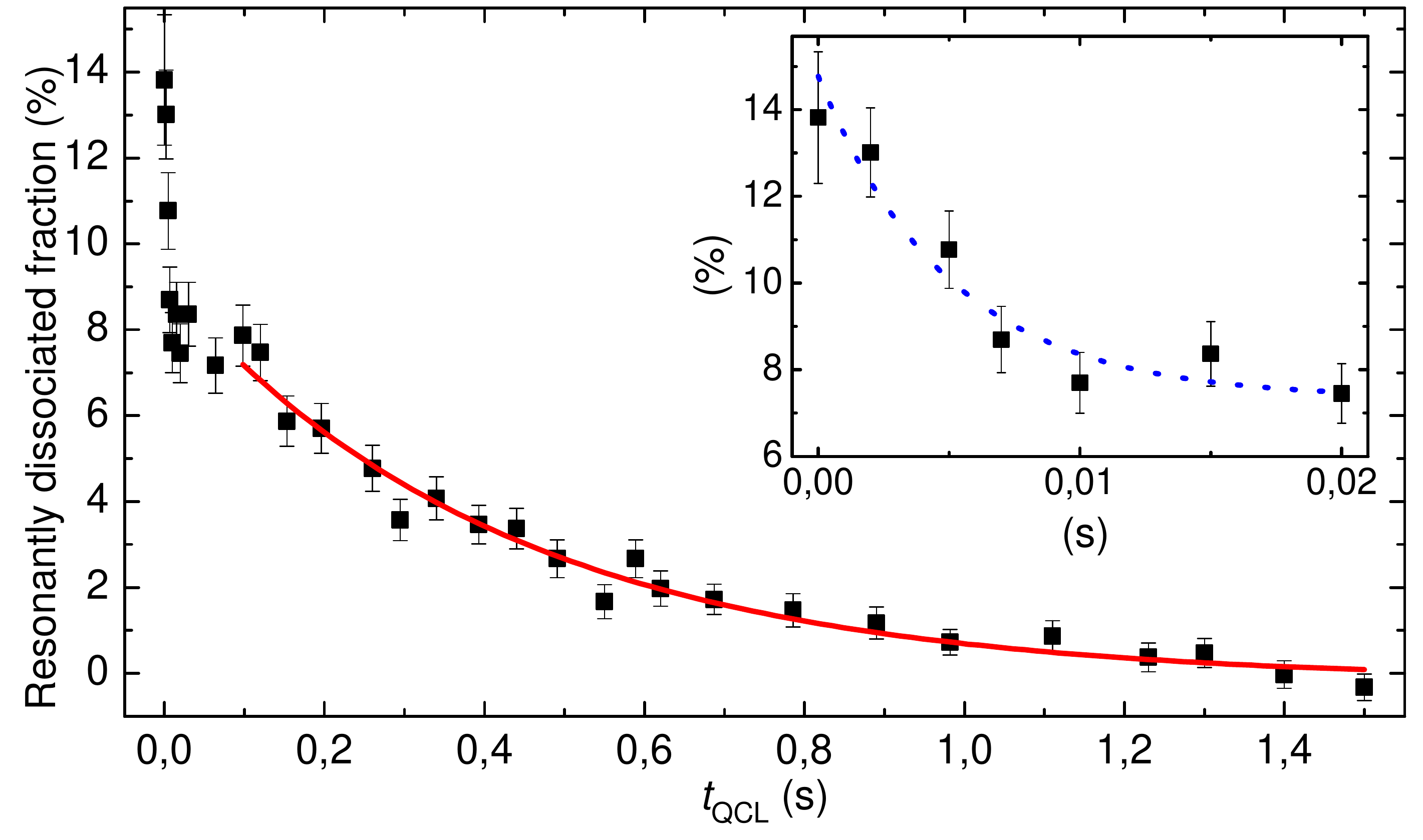}
\caption{Main figure: relative population of the \zt\, level of the \mgh\, molecules after application of the QCL for a time $t_{\textrm{QCL}}$. The solid red curve represents an exponential fit to the data points (black squares). Insert: close-up of the data for the first 20\,ms, including a exponential fit to this data selection (blue dotted line).}
\label{fig:p2lifetime}
\end{figure}

The employed QCL operates at the \mbox{6215.19\,nm} wavelength of the \zt-\oo\, transition \cite{Balfour1973}, and is referenced to an ammonia vapor absorption cell. The linearly polarized laser light has $\sim$30\,mW power in an approximately elliptic Gaussian beam profile of roughly \mbox{3.4$\times$1.6\,mm$^2$} (width at 1/$e^{2}$ intensity level) at the position of the ion crystal. To be sure to always address the \zt-\oo\, transition with the very narrow natural line width of only $2\pi\times$2.5\,Hz independently of long-term drift of the QCL center wavelength, this laser's line width is effectively broadened to \mbox{1.5\,GHz} (full-width at half-maximum, FWHM) through current modulation at 400\,kHz. A simple estimate under these conditions results in an effective Rabi frequency, $\Omega_\textrm{QCL}$, of $\sim$1000\,Hz, and hence confirms the validity of the assumption $\Gamma^{1,1}_{0,2}$, $\Gamma^{1,1}_{0,0} \ll \Omega_\textrm{QCL}$ made under the simplified decay model. Furthermore, with an uncompensated magnetic field at the position of the crystal of the order of the Earth's magnetic field ($\sim$0.5\,gauss), we expect the Larmor precession frequencies among the rotational magnetic sub-states to be of the order of 100\,Hz--1\,kHz \cite{townes1975}, which secures the last assumption $\Gamma_{B\textrm{-Mix.}} \gg \Gamma^{1,1}_{0,0}$ of the simplified model. Fig.\,\ref{fig:p2lifetime} shows the measured relative population of the \zt\, level as a function of the time $t_{\textrm{QCL}}$ that the \mgh\, ions were exposed to the QCL light. As expected, equilibration of the population between magnetic sub-states occurs rapidly and leads to equal population of the sub-states of the \zt\, and \oo\, levels (see Fig.\,\ref{fig:p2lifetime} inset). After 20\,ms, the population in the \zt\, level is reduced to a level of about 5/8 of the initial level indicating the establishment of equal population in all the sub-states. The measured rate for reaching this equilibrium of 199(69)\,s$^{-1}$ is in good agreement with that expected from the QCL parameters and the estimated magnetic field. 

After these initial fast dynamics, a slower exponential decay in the population is observed. The solid curve in Fig.\,\ref{fig:p2lifetime} represents an exponential fit to these data, and yields a decay rate $\Gamma^{\text{meas}}$ of $2.37(0.26)$\,s$^{-1}$. Using the simple decay model validated above, the decay rate $\Gamma^{1,1}_{0,0}$ can now be calculated to be $\Gamma^{1,1}_{0,0} =\frac{8}{3} \Gamma^{\text{meas}} = 6.32(0.69)$\,s$^{-1}$. 
Currently, the main uncertainty in the data points arises from uncertainties in the determination of the exact number of molecular ions from crystal sizes, the exact fractional population of the \oo\, level, non-resonant contribution to the REMPD process and the dissociation efficiency. One strategy to get around the first point would be to perform similar experiments but with single molecular ions \cite{PhysRevLett.100.243003,Hansen2012}. In any case, with improved laser stability, decay rate measurements with an accuracy at the few percent level should become feasible through the application of this new technique. 

The decay rate can be calculated from the known potential energy and dipole moment curves obtained by electronic structure methods \cite{jensen2007book}. We have performed a systematic investigation of the effects of basis-sets and electron correlation on the theoretically predicted decay rate. Electronic structure calculations were performed \cite{gaussian09} at the CCSD(T) level, where all electrons were correlated, using the uncontracted versions of the aug-cc-pCVXZ (X = D,T,Q,5) basis-sets \cite{prascher2011gaussian}. Convergence of the diffuse functions was achieved using a doubly-augmented cc-pCVQZ basis-set. Relativistic effects were estimated by the DKH2 formalism. Extrapolation to the basis-set limit was achieved using an exponential fit to (T,Q,5) data for the HF energy and an $L^{-3}$ fit to (Q,5) data for the correlation energy \cite{feller2011effectiveness}. Convergence of the CCSD(T) method with respect to increased electron correlation was achieved using the BD(TQ) method with the aug-pCVDZ and aug-pCVTZ basis-sets. Dipole moments relative to the COM system were calculated by a double numerical differentiation with respect to an external electric field of $\pm 10^{-4}$\,a.u. Analyses of the convergence of the theoretical decay rate with respect to basis set, electron correlation and relativistic effects, lead to a best estimate of 6.13(0.03)\,s$^{-1}$. This value is in excellent agreement with the experimental value, and in fair agreement with a previous theoretical estimate of 5.58\,s$^{-1}$ using a CI approach with a TZP quality basis set and an effective core potential for Mg \cite{Dulieu2009} (see Table\,\ref{tab:results}).

In summary, we have measured the decay of the $|\nu$=$1,J$=$1 \rangle_X$ towards the $|\nu$=$0,J$=$0 \rangle_X$ level in \mgh\ by saturated laser excitation of the $|\nu$=$0,J$=$2 \rangle_X$-$|\nu$=$1,J$=$1 \rangle_X$ transition followed by a 1+1 REMPD process which selectively dissociates only those molecules in the $|\nu$=$0,J$=$2 \rangle_X$ level. The cryogenic trap provided a low intensity BBR environment and ultra-high vacuum conditions thus yielding experimental conditions under which simple models can be constructed and used for analysis. 
\begin{table}[t!]\centering
\caption{Experimental and theoretical values for the \oo$\rightarrow$\zz\, decay rate in MgH$^+$.}
\begin{tabular}{llllll}
\hline \hline
Experiment\footnotemark[1] && Theory\footnotemark[1] \footnotemark[2] && Theory\footnotemark[2] \cite{Dulieu2009}& \rule{0pt}{2.6ex} \tabularnewline\hline
6.32(0.69)\,s$^{-1}$ && 6.13(0.03)\,s$^{-1}$ && 5.58\,s$^{-1}$& \rule{0pt}{2.6ex} \tabularnewline\hline\hline
\end{tabular}\footnotetext[1]{This work.} \footnotetext[2]{Calculated dipole moments were used to calculate decay rates employing the program {\sc LEVEL} \cite{Level}.}
\label{tab:results}
\end{table}

This work describing the measurement of a decay rate on laser-prepared molecular ions bridges the long-standing gap between such measurements and those on trapped cold \emph{neutral} molecules \cite{Meijer2005,Hoekstra2008,Doyle2008}. The method can readily be extended to other molecular ions of particular astrophysical importance such as CH$^+$ \cite{ornellas:1296,Wolf1998}, HD$^+$ \cite{Schiller2010}, OH$^+$ \cite{Vogelius2004}, and NH$^+$ \cite{Vogelius2004} with potential decay rate measurements at a few percent accuracy level, and hence provides an interesting alternative to other well-established techniques \cite{Church1993,calamai1994,jullien1994,Mauclaire1995,Wester1999,Amitay1999,hellberg2003}. The weak influence of the trapping environment on the molecular dynamics furthermore indicates that our cryogenic trap should as well be suited for ultra-high resolution molecular spectroscopy with the aim of testing fundamental physics \cite{PhysRevLett.100.023003,isaev2010laser,hudson2011improved,PhysRevA.86.050501}, for quantum state preparation and manipulation of molecular ions \cite{Shuman2010,Schiller2010,DrewsenNature2010,kahra2012}, and eventually for studies of ultra-cold chemistry \cite{softley2008,junye2009}.  
The MPIK mechanical workshops, and in particular its excellent apprentice workshop, have been of crucial importance for the construction of CryPTEx. OOV, MS, LK, and AW acknowledge funding from STSM travel grants from COST-Action IOTA. FJ was supported by grants from the Danish Center for Scientific Computation and the Danish Natural Science Research Council. The AU team furthermore appreciates significant support through the Danish National Research Foundation Center for Quantum Optics - QUANTOP, The Danish Agency for Science, Technology and Innovation, the Carlsberg Foundation as well as the Lundbeck Foundation. 

\bibliography{lifetimebib_15MAY2013_submission}
\end{document}